# Effects of turbulence spreading and symmetry breaking on edge shear flow during sawtooth cycles in J-TEXT tokamak[*]

DING Xiaoguan[1], ZHAO Kaijun[1, *], XIE Yaoyu[1], CHEN Zhipeng[2], CHEN Zhongyong[2], YANG Zhoujun[2], GAO Li[2], DING Yonghua[2], WEN Siyu[1], HU Yingxin[1]

1. School of Nuclear Science and Engineer, East China University of Technology, Nanchang 330013, China
2. State Key Laboratory of Advanced Electromagnetic Technology, International Joint Research Laboratory of Magnetic Confinement Fusion and Plasma Physics, School of Electrical and Electronic Engineering, Huazhong University of Science and Technology, Wuhan 430074, China

**Abstract**

The effect of sawteeth on plasma performance and transport in the plasma of tokamak is an important problem in the fusion field. Sawtooth oscillations can trigger off heat and turbulence pulses that propagate into the edge plasma, and thus enhancing the edge shear flow and inducing a transition from low confinement mode to high confinement mode. The influences of turbulence spreading and symmetry breaking on edge shear flow with sawtooth crashes are observed in the J-TEXT tokamak. The edge plasma turbulence and shear flow are measured using a fast reciprocating electrostatic probe array. The experimental data are analyzed using some methods such as conditional averaged and probability distribution function. After sawtooth crashes, the heat and turbulence pulses in the core propagate to the edge, with the turbulence pulse being faster than the heat pulse. The attached figures (a)–(e) show the core electron temperature, and the edge electron temperature, turbulence intensity, turbulence drive and spreading rates, Reynolds stress and its gradient, and shearing rates, respectively. After sawtooth crashes, the edge electron temperature increases and the edge turbulence is enhanced, with turbulence preceding temperature. The enhanced edge turbulence is mainly composed of two parts:

the turbulence driven by local gradient and the turbulence spreading from core to edge. The development of the estimated turbulence spreading rate is prior to that of the turbulence driving rate. The increase in the turbulence intensity can cause the turbulent Reynold stress and its gradient to increase, thereby enhancing shear flows and radial electric fields. Turbulence spreading leads the edge Reynolds stresses to develop and the shear flow to be faster than edge electron temperature. The Reynolds stress arises from the symmetry breaking of the turbulence wave number spectrum. After sawtooth collapses, the joint probability density function of radial wave number and poloidal wave number of turbulence intensity becomes highly skewed and anisotropic, exhibiting strong asymmetry, which can be seen in attached figures (f) and (g). The development of turbulence spreading flux at the edge is also prior to the particle flux driven by turbulence, indicating that turbulent energy transport is not simply accompanied by turbulent particle transport. These results show that the turbulence spreading and symmetry breaking can enhance turbulent Reynolds stress, thereby driving shear flows, after sawtooth has crashed.

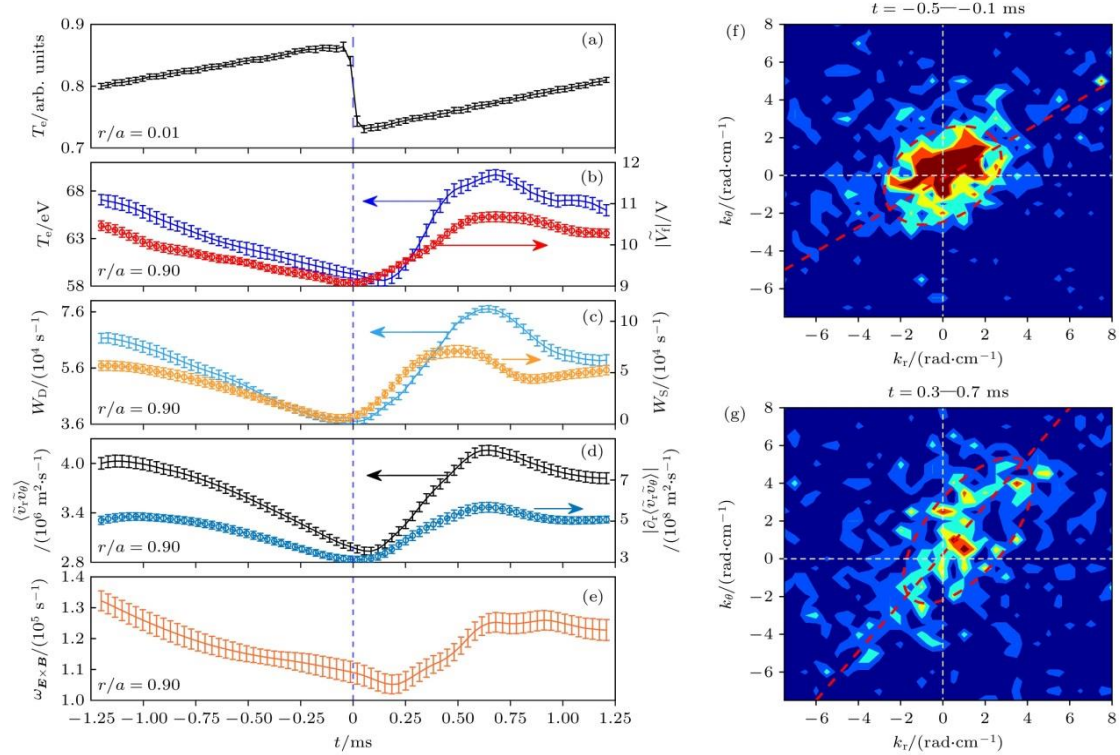

# 1. Introduction

Sawtooth oscillation, also known as sawtooth instability or sawtooth collapse, is a common macroscopic instability[1] in magnetically confined plasma. Since the discovery of[2] on ST tokamak in 1974, it has been an active research field in fusion community. Sawtooth oscillation is characterized by periodic rapid collapse and slow reconstruction of core temperature and density due to the disturbance of plasma core. At the same time, the particles and energy in the core are transported outward rapidly, which leads to the periodic sudden increase and slow decrease of the temperature and density of the external plasma, showing an anti-sawtooth shape.

At present, it has been observed on multiple devices such as ASDEX[3], JFT-2M[4], and TCV[5] that sawtooth can trigger the transition from L mode (low confinement mode, L mode) to H mode (high confinement mode, H mode). Research has shown that during the L-H transition, tokamak edge shear flow intensifies, turbulence is significantly suppressed, pressure gradient rapidly increases, and edge transport barrier forms. The plasma transport level decreases, and the confinement performance is greatly improved[6,7]. Shear flow can be generated by the self-organization of turbulence, and it suppresses turbulence through shear decorrelation while extracting energy from turbulent fluctuations[8].

Sawtooth oscillation can produce density, heat and turbulence pulses propagating from the core to the edge plasma, thus changing the edge density, temperature and turbulence to affect the plasma shear flow. Significant enhancement of edge plasma flow and turbulence after sawtooth collapse has been observed on the HL-2A[9,10] and J-TEXT[11] devices. Plasma turbulence can be either driven by localized gradients or spreading from turbulence at radial distances much larger than its correlation length. The effects of turbulence driving and spreading on the intensity and spatial distribution of edge turbulence were observed in the TJ-II[12] and J-TEXT[13-15]. However, how the turbulence spreading affects the time evolution of edge plasma turbulence and the generation mechanism of shear flow during sawtooth oscillation need to be further studied.

The effects of turbulence spreading and symmetry breaking on edge turbulence and shear flow during sawtooth oscillation are studied on the J-TEXT tokamak using a fast reciprocating electrostatic probe array. The results are of great significance for understanding plasma confinement and transport. Section 2 describes the experimental arrangement, Section 3 analyzes the effects of sawtooth oscillation on edge turbulence and shear flow, as well as the role of turbulence spreading and symmetry breaking on them, and Section 4 concludes the paper.

# 2. Experimental arrangement

2.1 Experimental setup

J-TEXT is a medium-sized tokamak device with a major radius of $R = 1.05\text{m}$ and a minor radius of $a = 0.255\text{m}$[16]. The experiment in this paper was carried out in a hydrogen plasma of ohmic discharge in the limiter configuration, employing a pulsed gas puffing fueling method. The discharge plasma parameters are plasma current $I_p = 190\text{ kA}$, line-integrated central density $n_e^{\text{LI}} = 1 - 1.7 \times 10^{19}\text{m}^{-2}$, toroidal magnetic field $B_t = 1.7\text{T}$ and edge safety factor $q_a = 2.7$.

The electron temperature of the core plasma was measured by a 24-channel electron cyclotron emission (ECE) system. The sampling rate of the ECE system is 1 MHz. Fast reciprocating electrostatic probe array was used to investigate the edge plasma parameters. The probe array is mounted on the top of the J-TEXT device and is movable in the radial direction 5 cm. The electrostatic probe array consists of three steps, each of which has four probes, and its structure is shown in the Fig. 1. All probes have a length of 3 mm and a diameter of 2 mm. The poloidal distance of probes 3 and 4, 7 and 8, and 11 and 12 is $d_\theta = 4\text{mm}$, which is much smaller than the poloidal correlation length 1—3cm of turbulence; the radial distance of adjacent steps is $d_r = 3\text{mm}$, which is much smaller than the radial correlation length 1—2cm of turbulence[17]. The positive bias $V_+$ is measured by probes 1, 5 and 9, the negative bias $V_-$ is measured by probes 2, 6 and 10, and the remaining probes measure the floating potential $V_f$. Through the three-step probe array, the plasma parameters such as local electron temperature, electron density, plasma potential, poloidal electric field and radial electric field can be observed. Electron temperature $T_e = (V_+ - V_f)/\ln 2$, ion saturation current $I_s = (V_+ - V_-)/R_S$, where $R_S$ is the shunt resistance, electron density $n_e = I_s/0.49eA_{\text{eff}}(T_e/m_i)^{1/2}$, where $A_{\text{eff}}$ is the effective current collection area, and $m_i$ is the ion mass. The plasma potential $V_p = V_f + \alpha T_e$ and $\alpha$ are taken as 2.5. The poloidal electric field $E_\theta = (V_f^{\theta_i} - V_f^{\theta_{i+1}})/d_\theta$, can be obtained from the poloidal gradient of the floating potential on the same step of the probe array, and the $i$ is 3, 7 or 11 probes. The $E_r = (V_p^{r_j} - V_p^{r_{j+1}})/d_r$ of the radial electric field can be obtained from the radial gradient of the plasma potential of two adjacent steps, and the $j$ is the first or second step. The radial $\boldsymbol{E} \times \boldsymbol{B}$ velocity is $V_r = E_\theta/B_t$, and the poloidal $\boldsymbol{E} \times \boldsymbol{B}$ velocity is $V_\theta = E_r/B_t$. The sampling rate of the probe is 2 MHz, corresponding to a Nyquist frequency of 1 MHz.

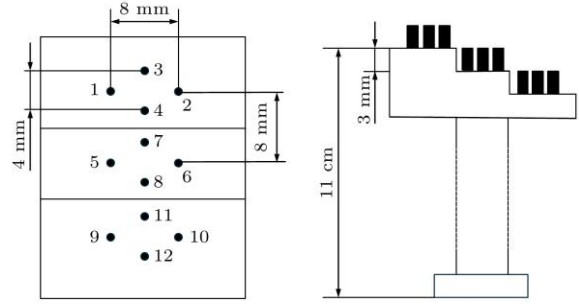

**Figure 1.** Structure of electrostatic probe array.

2.2 Analytical method

The conditional averaged method can identify the coherent structure in the signal and suppress the signal noise. When the conditional averaged method is used to analyze data, the reference signal needs to be selected first. In this experiment, the sawtooth collapse is the reference signal and is calculated from the time derivative of the ECE signal at $r/a = 0.01$. In each sawtooth period, the time at which the derivative is at a minimum is defined $t = 0$ ms. The average period of the sawtooth is 2.5 ms, and the $t = -1.25 - 1.25$ ms is selected as the time window to extract each signal sequence. The conditional average result of the signal can be obtained by superimposing and averaging all the extracted sequences. A total of 25 sawtooth data were extracted for conditional average analysis.

# 3. Experimental results and discussion

This analysis uses experimental data from shot #1063158. Fig. 2(a) —(c) represent the plasma current $I_\mathrm{p}$, the line integrated electron density $n_\mathrm{e}^\mathrm{LI}$, the toroidal magnetic field $B_\mathrm{t}$, respectively. The Fig. 2(d) is the electron temperature at the $r/a = 0.01$ measured by the electron cyclotron radiation diagnostic system, in arbitrary units. At present, the electron cyclotron radiation diagnostic on the J-TEXT device is only calibrated relatively, and the actual temperature cannot be given. Fig. 2(e) and Fig. 2(f) are the time evolution of the floating potential and the probe displacement at $r/a = 0.88$, respectively, where $\Delta r$ represents the distance between the last closed flux surface (LCFS) and the probe measurement position, and the negative sign represents inside the LCFS. At $t = 270$ ms, the probe begins to insert into the plasma and stops moving at $t = 340$ ms, staying about 3 cm inside the last closed magnetic surface, with a floating potential of about $-170$ V. The shaded time period is $t = 367 - 423$ ms, and the probe data in this time period is analyzed in detail below.

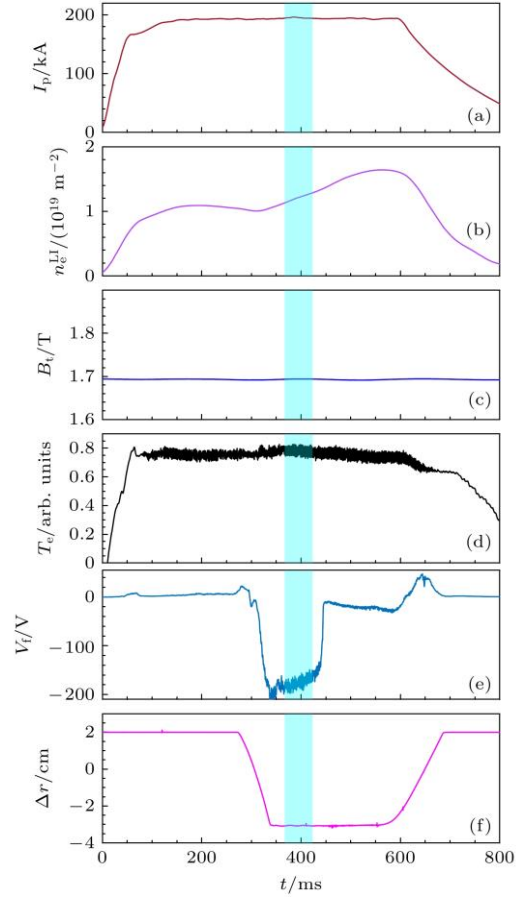

**Figure 2.** Plasma discharge parameters: (a) Plasma current; (b) line integrated electron density; (c) toroidal magnetic field; (d) electron temperature at $r/a = 0.01$; (e) floating potential at $r/a = 0.88$; (f) probe positions.

3.1 Plasma turbulence spreading

After the sawtooth collapse, the thermal and turbulent pulses propagate outward from the inversion radius to the edge plasma. The electron temperature and turbulent relative intensity at different radial positions are given by Fig. 3(a) -(h), respectively. The electron temperature fluctuation $\tilde{T}_e$ here is derived from the ECE signal, and the corresponding frequency band is 10-50 kHz. The method of extracting turbulence signal from ECE signal is introduced in detail in[11,18]. The error bars in the figure are given by the standard deviation of the conditionally averaged corresponding signal, the blue dotted line represents the $t = 0$ms, which corresponds to the sawtooth collapse time at the $r/a = 0.01$, the green solid line and the dotted line connect the peak values of the relative turbulence intensity and the electron temperature, respectively, and the arrow represents the propagation direction. As can be seen from the Fig. 3, the reversal radius is around $r/a = 0.39$. Within the reversal radius, the sawtooth collapse and the turbulence pulse propagate inward, and the relative turbulence intensity at $r/a = 0.29, 0.19, 0.10$ and $0.01$ increases by 257%, 378%, 442% and 381%, respectively. The electron temperature at $r/a = 0.5, 0.61$ and $0.72$ increases by 18%, 16% and 16%, respectively, and the relative turbulence intensity increases by 219%, 122% and 90%, respectively. The turbulent pulse propagates faster than the thermal pulse, with velocities of 0.6 km/s and 0.2 km/s, respectively.

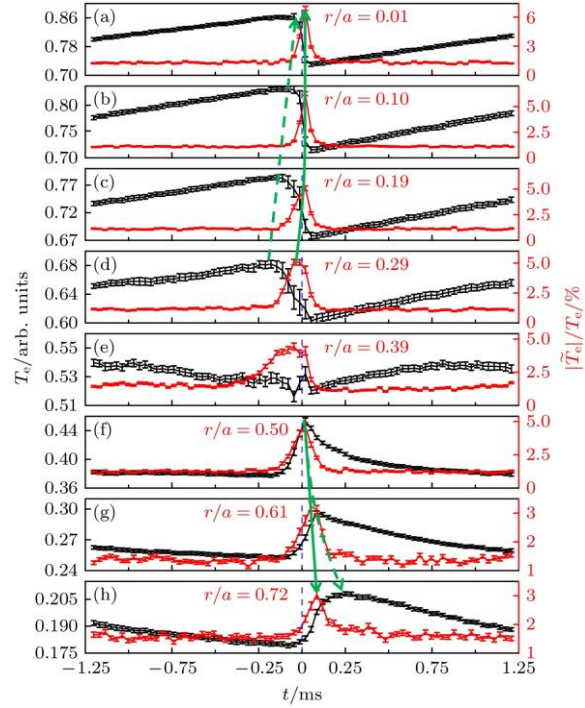

**Figure 3.** Electron temperature and relative intensities of turbulence at various radial positions, respectively.

The Fig. 4(a) -(d) gives the electron temperature at the $r/a = 0.01$, the electron temperature at the $r/a = 0.90$, the floating potential, and the turbulence intensity, respectively. The edge turbulence intensity is given by the absolute fluctuation $|\tilde{V}_f|$ of the floating potential, and the fluctuation is obtained by filtering the edge turbulence frequency band (30-150 kHz) of the corresponding signal. According to the Fig. 4(a) -(d), the edge electron temperature increases by 20% to the maximum and the floating potential decreases by 20% to the minimum 0.7 ms after the sawtooth collapse. The edge turbulence intensity increases by 20% to a maximum after 0.6 ms of sawtooth collapse. The Fig. 4(g) gives the Lissajous figure of the electron temperature and turbulence intensity at $r/a = 0.90$, where the red arrow indicates the direction of rotation of the trajectory and the red dot indicates the time of sawtooth collapse at $r/a = 0.01$. As shown in Fig. 4(g), the trajectory rotates clockwise as a whole, and after the sawtooth collapse, the edge turbulence develops faster than the electron temperature and reaches its maximum first. This indicates that the turbulent pulse propagating from the core to the edge is faster than the thermal pulse after the collapse of the sawtooth, and the delay time of the thermal pulse is about 0.1 ms.

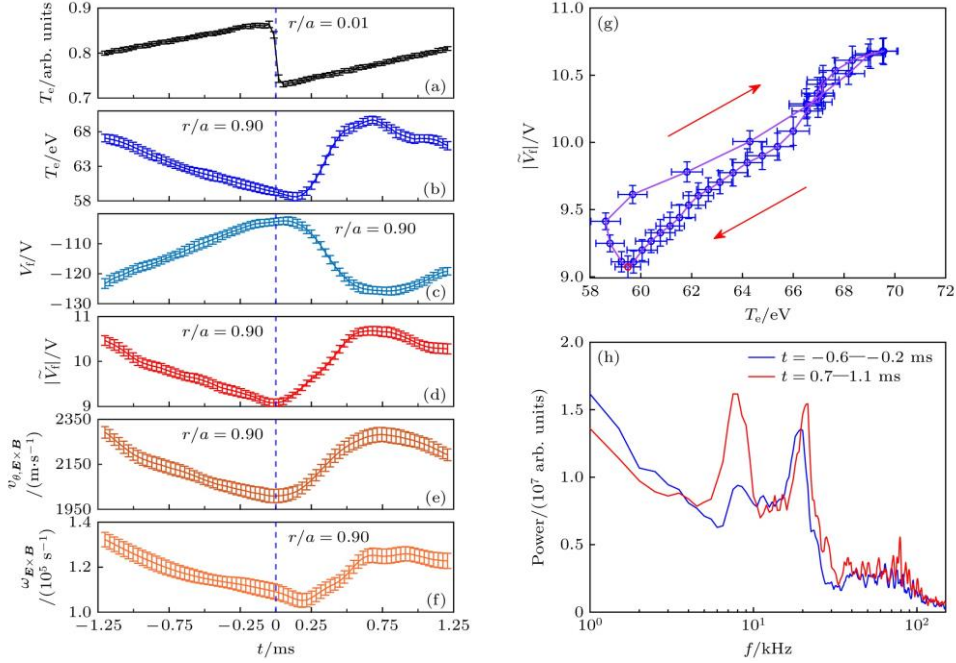

**Figure 4.** (a) Electron temperature at $r/a = 0.01$; (b) electron temperature at $r/a = 0.90$; (c) floating potential at $r/a = 0.90$; (d) turbulence intensity at $r/a = 0.90$; (e) $E \times B$ poloidal velocity at $r/a = 0.90$; (f) shearing rate at $r/a = 0.90$; (g) trajectory of electron temperature and turbulence intensity at $r/a = 0.90$; (h) auto-power spectra of the floating potential before and after sawtooth collapse at $r/a = 0.90$.

The Fig. 4(h) gives the auto-power spectrum of the floating potential at the $r/a = 0.90$ before and after the sawtooth collapse. It can be seen from the figure that the $f = 30-150$ kHz is the turbulent frequency band, the geodesic acoustic mode (GAM) has a frequency of 18 kHz, and the spectral peak at 8 kHz is the tearing mode. After the sawtooth collapse, the turbulence is significantly enhanced, the GAM driven by the turbulence is also enhanced, and the peak frequency increases from 18 kHz to 21 kHz, and the tearing mode instability also increases.

Turbulence can self-organize to produce shear flow through nonlinear interaction. The Fig. 4(e) and Fig. 4(f) give the poloidal velocity of the $E \times B$ and its shear rate at the $r/a = 0.90$, respectively. It can be seen that after the sawtooth collapse, as the turbulence intensity increases, the poloidal velocity of the edge $E \times B$ increases from 2 km/s to 2.3 km/s, an increase of about 15%, and the poloidal $E \times B$ flow shear rate increases from $1.0 \times 10^5$ s$^{-1}$ to $1.2 \times 10^5$ s$^{-1}$. In general, stronger shear flow helps to improve confinement.

After sawtooth collapse, the edge-enhanced turbulence mainly consists of two parts, namely, the turbulence driven by the local gradient at the edge and the turbulence spreading from the core to the edge. Through the study of the evolution of turbulent energy by Manz et al.[19], we can identify these two types of turbulence with different sources. Neglecting the cross-field coupling and damping and the effect of the background flow, and considering in-plane incompressible turbulence perpendicular to the magnetic field, the turbulent energy evolution equation can be derived from the radial part of the continuity equation:

$$\frac{1}{2}\frac{\partial}{\partial t}\langle \tilde{n}^2 \rangle = -\left\langle \frac{\partial n}{\partial r} \right\rangle \langle \tilde{v}_r \tilde{n} \rangle - \frac{1}{2}\frac{\partial}{\partial r}\langle \tilde{v}_r \tilde{n}^2 \rangle. \tag{1}$$

Here, $\langle \cdots \rangle$ represents an average over a time interval much longer than the turbulence correlation time. The first term on the left side of the equation represents the time evolution of the turbulent free energy, and the $\tilde{n}$ is the density fluctuation. The first term on the right hand side of the equation is related to the driving of turbulence by local gradients, $\langle \tilde{v}_r \tilde{n} \rangle$ represents particle transport, and $\tilde{v}_r$ is the radial velocity fluctuation. The second term on the right hand side of the equation is the nonlocal nonlinear turbulent spreading. The turbulence driving rate driven by the local density gradient is obtained by normalizing with the local turbulence amplitude $\langle \tilde{n}^2 \rangle$:

$$\omega_D = -\frac{2\langle \frac{\partial n}{\partial r} \rangle \langle \tilde{v}_r \tilde{n} \rangle}{\langle \tilde{n}^2 \rangle}. \tag{2}$$

The turbulent spreading rate is then expressed as

$$\omega_S = -\frac{\frac{\partial}{\partial r}\langle \tilde{v}_r \tilde{n}^2 \rangle}{\langle \tilde{n}^2 \rangle}. \tag{3}$$

The Fig. 5(a) -(d) gives the electron temperature at the $r/a = 0.01$, the turbulence intensity at the $r/a = 0.90$, the turbulence driving rate, and the turbulence spreading rate, respectively. According to the Fig. 5, the turbulence driving rate reaches a maximum of $7.6 \times 10^4$ s$^{-1}$ after 0.7 ms of sawtooth collapse. The turbulence spreading rate reaches a maximum of $7.0 \times 10^4$ s$^{-1}$ after 0.6 ms of sawtooth collapse. The turbulence spreading rate increases about 0.1 ms faster than the driving rate. This result is consistent with the result that the turbulent pulse is faster than the thermal pulse, which indicates that the edge turbulent pulse is faster than the thermal pulse due to turbulent spreading. It can be seen that after 0.6 ms of sawtooth collapse, the turbulent pulse propagating outward from the inversion radius reaches the edge earlier than the thermal pulse rate, which enhances the edge turbulence. Turbulence spreading comes from the nonlinear interaction of turbulence. Generally, the increase of turbulence intensity will lead to the enhancement of nonlinear interaction of turbulence, and thus the enhancement of turbulence spreading. When the heat pulse reaches the edge 0.7 ms after the sawtooth collapse, it affects the local gradient distribution at the edge, and the turbulent driving rate increases with the steepening of the local gradient. This is because the linear driving of turbulence comes mainly from the gradient. In addition, the maximum values of turbulence driving rate and turbulence spreading rate are similar after the collapse of the sawtooth, which indicates that both the turbulence driven by the local gradient at the edge and the turbulence spreading from the core to the edge may play an important role in the enhancement of turbulence at the edge.

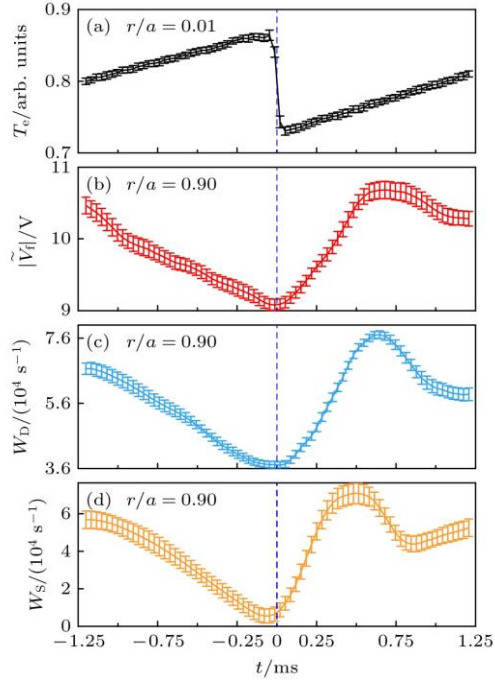

**Figure 5.** (a) Electron temperature at $r/a = 0.01$; (b) turbulence intensity at $r/a = 0.90$; (c) turbulence drive rates at $r/a = 0.90$; (d) turbulence spreading rates at $r/a = 0.90$.

Turbulent spreading refers to the spatial spreading of turbulent intensity or energy due to nonlinear interactions. Turbulent energy consists of turbulent internal energy and kinetic energy. For the turbulent internal energy, $\langle \tilde{v}_r \tilde{n}^2 \rangle \langle C_s^2 \rangle / (2 \langle n \rangle^2)$ is the associated flux, and $C_s$ is the ion sound speed. For turbulent kinetic energy, $\langle \tilde{v}_r \tilde{v}_\perp^2 \rangle / 2$ is the correlation flux and $\tilde{v}_\perp$ is the vertical velocity fluctuation. In general, the kinetic energy flux observed in the experiment is about 2 orders of magnitude smaller than the internal energy flux[20,21], so it can be ignored in this calculation. For simplification, the turbulent propagation flux can be calculated from the $\langle \tilde{v}_r \tilde{n}^2 \rangle / 2$. This calculation method has also been applied in theory and simulation[22,23]. In order to further understand the characteristics of turbulent spreading, the turbulent mean jet velocity is calculated $V_I = \langle \tilde{v}_r \tilde{n}^2 \rangle / \langle \tilde{n}^2 \rangle$. The mean jet velocity of turbulent spreading represents the characteristic velocity of turbulent internal energy spreading in space. It can be seen from Fig. 6(b) and Fig. 6(d) that the turbulent spreading flux and the mean jet velocity reach the maximum after 0. 6 ms of sawtooth collapse, which is consistent with the change of turbulent spreading rate in Fig. 5(d). In order to compare the turbulent spreading flux with the turbulent driving flux, the turbulent particle flux $\langle \tilde{v}_r \tilde{n} \rangle$ and the particle transport velocity are also calculated $V_T = \langle \tilde{v}_r \tilde{n} \rangle / \langle \tilde{n}^2 \rangle^{1/2}$. The particle transport velocity can be regarded as the characteristic velocity of turbulent particle flux transport. It can be seen from Fig. 6(c) and Fig. 6(e) that the turbulent particle flux and particle transport velocity reach their maximum values after 0.7 ms of sawtooth collapse, which is consistent with the change of turbulent driving rate in Fig. 5(c). The analysis shows that the time for the turbulent spreading flux to reach its maximum after sawtooth collapse is faster than the turbulent particle flux by about 0.1 ms. This shows that turbulent energy transport is not simply accompanied by turbulent particle transport. In addition, the average jet velocity of particle transport and turbulent spreading during sawtooth oscillation is estimated to be 990 m/s and 150 m/s, respectively, which is similar to the results reported by Long Ting et al[14].

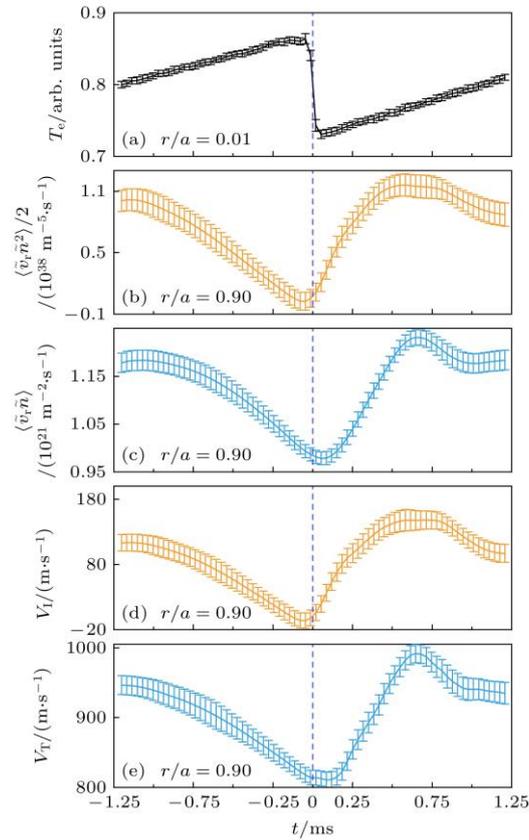

**Figure 6.** (a) Electron temperature at $r/a = 0.01$; (b) turbulence spreading flux at $r/a = 0.90$; (c) turbulence particle flux at $r/a = 0.90$; (d) mean jet velocity of turbulence spreading at $r/a = 0.90$; (e) particle transport velocity at $r/a = 0.90$.

3.2 Effect of Sawtooth Oscillation on Radial Electric Field

The $\boldsymbol{E} \times \boldsymbol{B}$ flow originates from the $\boldsymbol{E} \times \boldsymbol{B}$ drift caused by the radial electric field $E_r$ and the toroidal magnetic field $B_t$. The radial electric field can be decomposed into three[24] through the radial force balance equation:

$$E_r = \nabla P_i / e n_i Z_i - v_\theta B_t + v_\phi B_\theta. \tag{4}$$

The first term on the right side of the equation represents the contribution of the mean flow driven by the ion pressure gradient to the radial electric field, and $\nabla P_i, n_i$, and $Z_i$ are the ion pressure gradient, ion density, and nuclear charge, respectively. In the edge plasma, assuming that the $T_e \approx T_i$ and the plasma quasi-neutrality condition are $n_e \approx n_i$ get $P_e \approx P_i$. The second term on the right hand side of the equation represents the contribution of the poloidal flow to the radial electric field, $v_\theta$ is the poloidal velocity, and $B_t$ is the toroidal magnetic field strength. The third term on the right hand side of the equation represents the contribution of the toroidal flow to the radial electric field, $v_\phi$ is the toroidal velocity, and $B_\theta$ is the poloidal magnetic field strength. In tokamak, the poloidal magnetic field is generally one order of magnitude smaller than the toroidal magnetic field, and the edge toroidal flow velocity and the poloidal flow velocity are close to[25] during L-mode discharge on J-TEXT. Therefore, the contribution of the toroidal flow to the radial electric field can be ignored by considering the poloidal magnetic field and the toroidal flow velocity. The radial force equilibrium equation can be simplified to $E_r = \nabla P_e / e n_e Z_i - v_\theta B_t$. The electron temperature at $r/a = 0.01$, the radial electric field strength at $r/a = 0.90$, the pressure gradient, and the contribution of the poloidal flow to the radial electric field are given by Fig. 7(a)-(d), respectively. From Fig. 7, it can be seen that the contribution of poloidal flow increases by 16% after 0.6 ms after the sawtooth collapse. After 0.7 ms of sawtooth collapse, the radial electric field intensity increases by about 16%, and the contribution of the pressure gradient increases by 13%. The contribution of pressure gradient and poloidal flow to the radial electric field is 25% and 75%, respectively. The results show that the poloidal flow plays a major role in the radial electric field during the sawtooth oscillation, and the poloidal flow increases faster than the pressure gradient after the sawtooth collapse.

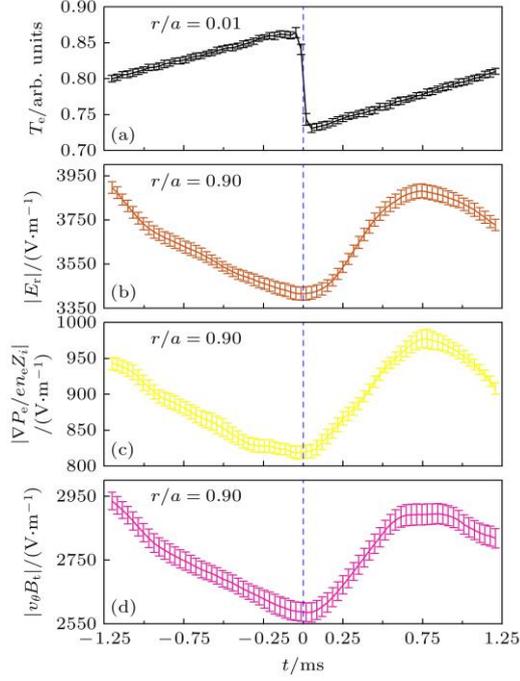

**Figure 7.** (a) Electron temperature at $r/a = 0.01$; (b) radial electric fields intensity at $r/a = 0.90$; (c) contributions of the pressure gradient at $r/a = 0.90$; (d) poloidal flows to the radial electric field at $r/a = 0.90$.

3.3 Effect of sawtooth oscillation on Reynolds stress in edge turbulence

In ohmic discharge, the poloidal flow primarily originates from turbulent Reynolds stress [26]. The Reynolds stress can be calculated from the $\langle \tilde{v}_r \tilde{v}_\theta \rangle$, and the $\tilde{v}_\theta$ is the poloidal velocity fluctuation. The radial gradient of the Reynolds stress can cause a redistribution of the poloidal momentum, resulting in a poloidal shear[27]. The power of energy conversion from turbulent flow through Reynolds stress to poloidal flow is called Reynolds power $P_{RS} = -\langle \tilde{v}_r \tilde{v}_\theta \rangle \partial \langle v_\theta \rangle / \partial r$. The Reynolds stress, the absolute value of the Reynolds stress gradient and the Reynolds power at the $r/a = 0.90$ are given by the Fig. 8(c) -(e), respectively. It can be seen that the Reynolds stress, the Reynolds stress gradient and the Reynolds power increase to the maximum after the sawtooth collapse for 0. 6 ms. Comparing Fig. 8(b),Fig. 8(d) and Fig. 8(f), the time for Reynolds stress gradient and shear flow to reach the maximum after sawtooth collapse is faster than the time for the edge electron temperature to reach the maximum, which is about 0.1 ms, and is consistent with the delay time of the edge turbulence spreading rate. This indicates that the turbulent spreading plays an important role in the edge Reynolds stress and its gradient, thus affecting the edge poloidal shear flow.

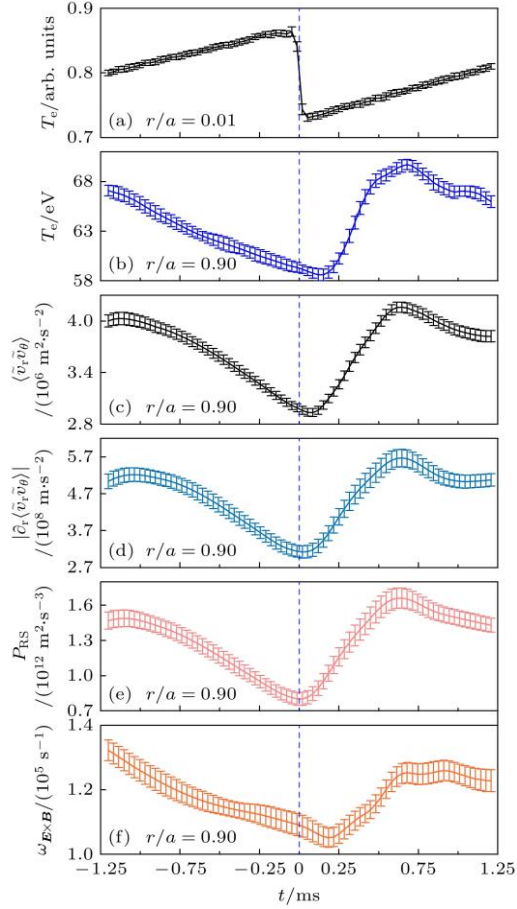

**Figure 8.** (a) Electron temperature at $r/a = 0.01$; (b) electron temperature at $r/a = 0.90$, (c) Reynolds stress at $r/a = 0.90$; (d) gradient of Reynolds stress at $r/a = 0.90$; (e) Reynolds power at $r/a = 0.90$; (f) shearing rate at $r/a = 0.90$.

Turbulent Reynolds stress originates from the symmetry breaking of the wavenumber spectrum of turbulence[28]. Reynolds stress $\langle \tilde{v}_r \tilde{v}_\theta \rangle \propto \langle k_\theta k_r \rangle |\tilde{\phi}|^2/B^2$, $k_r$ and $k_\theta$ are the radial and poloidal wave numbers, respectively, $\tilde{\phi}$ is the potential fluctuation, and $B$ is the magnetic flux density. It can be seen from the formula that when the geometric structure of the turbulent vortex is symmetrical, that is, $\langle +k_\theta \rangle = \langle -k_\theta \rangle$, the Reynolds stress produced is 0. When the geometric symmetry of the turbulent vortex is broken, that is, $\langle +k_\theta \rangle \neq \langle -k_\theta \rangle$, $\langle k_r k_\theta \rangle = \sum_k k_r k_\theta |\tilde{\phi}_k|^2 / \sum_k |\tilde{\phi}_k|^2 \neq 0$, a non-zero Reynolds stress can be produced. In order to study the mechanism of Reynolds stress, the joint probability density function $P(k_r, k_\theta) = N_{k_r k_\theta}/N_0$, $N_{k_r k_\theta}$ of turbulent $k_r$ and $k_\theta$ is calculated. is the number of events falling in the interval of $(k_r, k_r + \Delta k_r)$ and $(k_\theta, k_\theta + \Delta k_\theta)$, and $N_0$ is the total number of events. As shown in Fig. 9(a) and Fig. 9(b), before the collapse of the sawtooth, the $P(k_r, k_\theta)$ is elongated in the radial direction and inclined at an angle of 30 °, which is closer to a symmetrical distribution. After the sawtooth collapse, the main distribution range of $k_\theta$ changes from $-2\text{—}2\,\text{rad}\cdot\text{cm}^{-1}$ to $-2\text{—}4\,\text{rad}\cdot\text{cm}^{-1}$, and $P(k_r, k_\theta)$ extends along the polar direction with an inclination angle of 50 °, mainly distributed in the first and third quadrants, and the spectrum as a whole moves to the positive $k_r$ direction. The $P(k_r, k_\theta)$ becomes highly tilted and anisotropic, exhibiting a strong asymmetry. This is consistent with the result that turbulence produces stronger Reynolds stress after sawtooth collapse.

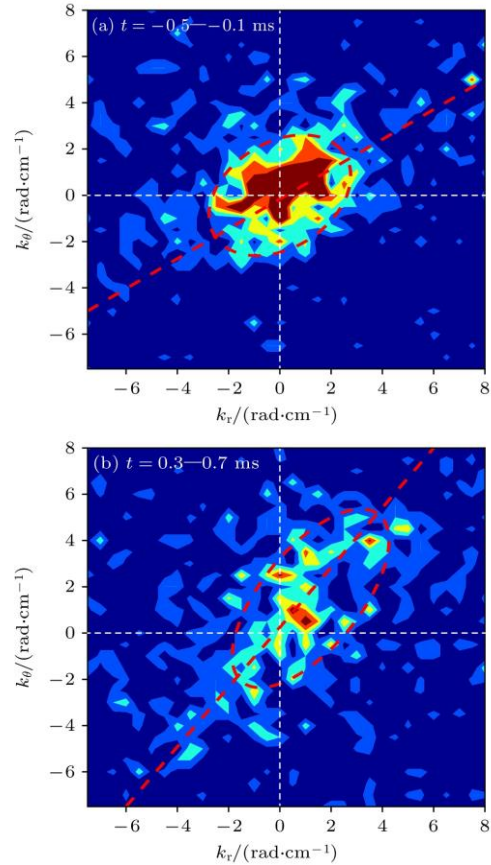

**Figure 9.** Joint probability density function of radial and poloidal wave numbers of turbulence intensity before (a) and after (b) sawtooth collapse at $r/a = 0.90$.

## 4. Summary

The effect of sawtooth on the confinement performance and transport level of magnetically confined edge plasma has been an important issue in the field of fusion. Previous studies have focused on the propagation of temperature, density, and gradients generated by serrations. In fact, sawtooth can also induce turbulent pulses to propagate to the edge, affecting edge plasma turbulence and shear flow. In this paper, the effects of turbulence spreading and symmetry breaking on edge turbulence and shear flow during sawtooth oscillation are observed for the first time by using a fast reciprocating electrostatic probe array. The experimental results are as follows.

1) During the sawtooth oscillation, the thermal pulse and the turbulent pulse propagate from the core to the edge plasma, and the propagation of the turbulent pulse at the edge is faster than that of the thermal pulse. After sawtooth collapse, the enhancement of edge turbulence is mainly composed of the turbulence driven by the local gradient at the edge and the turbulent pulse propagating from the core to the edge. The important reason why the turbulence develops faster than the thermal pulse at the edge is that the turbulence spreading is faster than the driving. The turbulence spreading flux also develops faster than the turbulence-driven particle flux at the edge, indicating that the transport of turbulent energy is not simply accompanied by turbulent particle transport.

2) After the collapse of the sawtooth, the edge-enhanced turbulence contributes to the formation of the shear flow. The development of the edge radial electric field is mainly contributed by the poloidal flow. The poloidal flow is mainly generated by the turbulent Reynolds stress. After the sawtooth collapse, the turbulent Reynolds stress and its gradient are enhanced, and their development is faster than that of the thermal pulse, indicating that the turbulence spreading plays an important role in it. After the sawtooth collapse, the turbulent joint probability density spectrum shows a strong asymmetry, which indicates that the symmetry breaking of turbulence after the sawtooth collapse can produce Reynolds stress. The results illustrate the effects of turbulence spreading and symmetry breaking on the edge plasma flow and turbulence. This study is of great significance for understanding plasma confinement and transport, and provides a reference for the design and operation of future fusion reactors.